\documentstyle[12pt,a4]{article}

\newcommand{\nc}[2]{\newcommand{#1}{#2}}
\newcommand{\ncx}[3]{\newcommand{#1}[#2]{#3}}
\ncx{\pr}{1}{#1^{\prime}}
\nc{\nl}{\newline}
\nc{\np}{\newpage}
\nc{\nit}{\noindent}
\nc{\be}{\begin{equation}}
\nc{\ee}{\end{equation}}
\nc{\ba}{\begin{array}}
\nc{\ea}{\end{array}}
\nc{\bea}{\begin{eqnarray}}
\nc{\eea}{\end{eqnarray}}
\nc{\nb}{\nonumber}
\nc{\dsp}{\displaystyle}
\nc{\bit}{\bibitem}
\nc{\ct}{\cite}
\ncx{\dd}{2}{\frac{\partial #1}{\partial #2}}
\nc{\pl}{\partial}
\nc{\dg}{\dagger}
\nc{\cH}{{\cal H}}
\nc{\cL}{{\cal L}}
\nc{\cD}{{\cal D}}
\nc{\cF}{{\cal F}}
\nc{\cG}{{\cal G}}
\nc{\cJ}{{\cal J}}
\nc{\cQ}{{\cal Q}}
\nc{\tB}{\tilde{B}}
\nc{\tD}{\tilde{D}}
\nc{\tH}{\tilde{H}}
\nc{\tK}{\tilde{K}}
\nc{\tR}{\tilde{R}}
\nc{\tZ}{\tilde{Z}}
\nc{\tg}{\tilde{g}}
\nc{\tog}{\tilde{\og}}
\nc{\tGam}{\tilde{\Gam}}
\nc{\tPi}{\tilde{\Pi}}
\nc{\tcD}{\tilde{\cD}}
\nc{\tcQ}{\tilde{\cQ}}
\nc{\ag}{\alpha}
\nc{\bg}{\beta}
\nc{\gam}{\gamma}
\nc{\Gam}{\Gamma}
\nc{\bgm}{\bar{\gam}}
\nc{\del}{\delta}
\nc{\Del}{\Delta}
\nc{\eps}{\epsilon}
\nc{\ve}{\varepsilon}
\nc{\zg}{\zeta}
\nc{\th}{\theta}
\nc{\vt}{\vartheta}
\nc{\Th}{\Theta}
\nc{\kg}{\kappa}
\nc{\lb}{\lambda}
\nc{\Lb}{\Lambda}
\nc{\ps}{\psi}
\nc{\Ps}{\Psi}
\nc{\sg}{\sigma}
\nc{\spr}{\pr{\sg}}
\nc{\Sg}{\Sigma}
\nc{\rg}{\rho}
\nc{\fg}{\phi}
\nc{\Fg}{\Phi}
\nc{\vf}{\varphi}
\nc{\og}{\omega}
\nc{\Og}{\Omega}
\nc{\Kq}{\mbox{$K(\vec{q},t|\pr{\vec{q}\,},\pr{t})$ }}
\nc{\Kp}{\mbox{$K(\vec{q},t|\pr{\vec{p}\,},\pr{t})$ }}
\nc{\vq}{\mbox{$\vec{q}$}}
\nc{\qp}{\mbox{$\pr{\vec{q}\,}$}}
\nc{\vp}{\mbox{$\vec{p}$}}
\nc{\va}{\mbox{$\vec{a}$}}
\nc{\vb}{\mbox{$\vec{b}$}}
\nc{\Ztwo}{\mbox{\sf Z}_{2}}
\nc{\Tr}{\mbox{Tr}}
\nc{\lh}{\left(}
\nc{\rh}{\right)}
\nc{\ld}{\left.}
\nc{\rd}{\right.}
\nc{\nil}{\emptyset}
\nc{\bor}{\overline}
\nc{\ha}{\hat{a}}
\nc{\da}{\hat{a}^{\dg}}
\nc{\hb}{\hat{b}}
\nc{\db}{\hat{b}^{\dg}}
\nc{\hN}{\hat{N}}
\ncx{\abs}{1}{\left| #1 \right|}
\nc{\vs}{\vspace{2ex}}
\nc{\vvs}{\vspace{3ex}}

\begin{document}

\pagestyle{empty}

\begin{flushright}
NIKHEF/95-067 \\
hep-th/9512068
\end{flushright}
\vspace{3ex}

\begin{center}
{\Large {\bf Fermions and world-line supersymmetry$^*$}} \\
\vspace{5ex}

{\large J.W.\ van Holten } \\
\vspace{3ex}

{\large NIKHEF/FOM, Amsterdam NL } \\
\vspace{7ex}

December 11, 1995
\vspace{9ex}

{\small
{\bf Abstract}}
\end{center}
\vs

\nit
{\small
The world-line path-integral representation of fermion propagators is
discussed. Particular attention is paid to the representation of $\gamma_5$,
which is connected to the realization of manifest  world-line supersymmetry.
}
\vfill

\nit
{\footnotesize
$\overline{ ^* \mbox{Contribution to Proceedings of 29th Int.\/}}$
            Symposium on the Theory of Elementary Particles \nl Buckow (1995)
}

\newpage
\pagestyle{plain}
\pagenumbering{arabic}

%
%
%
%

\nit
{\bf Introduction}
\vs

The path-integral representation of the evolution operator of dynamical
quantum systems provides a direct link to the associated (pseudo-)classical
system. This classical theory can suggest symmetries of the quantum
theory (see for example \ct{RvH1}) and reformulations taking advantage of
various classical descriptions related by co-ordinate transformations (or
canonical transformations in the Hamiltonian formalism). As has been emphasized
by many authors \ct{pol}-\ct{jw1}, Green's functions in perturbative
quantum field theory can also be represented in terms of world-line
path integrals, providing additional insight into their properties, like
analyticity and causality. This formulation also provides new calculational
tools, some of them inspired by considering the point-particle limit of string
theory \ct{bern}.

As was realized long ago \ct{ber1}-\ct{bcl} a convenient description of
fermions by world-line path integrals requires the introduction of Grassmann
variables, and it was found that world-line supersymmetry imposes a structure
appropriate to the description of Dirac fermions. Applications of these ideas,
as for example the computation of fermion determinants in field theory,
require the inclusion of external background fields like scalars, vectors
or tensors \ct{bcl}-\ct{jw3}.

An important issue in the construction of fermion world-line path integrals
is the representation of $\gam_5$, as it appears in mass terms \ct{jw4},
axial couplings \ct{dhok,mon2} and anomalies \ct{alvG}. In practice there are
two approaches to this issue, which have led a more or less independent life
in the literature. The first one, motivated by supersymmetry, is to introduce
a Grassmann-odd variable $\ps_5$ in addition to the Grassmann-odd vector
variables $\ps_{\mu}$, with its own kinetic terms in the pseudo-classical
action \ct{ber1}-\ct{bcl}. The other one, motivated by the computation of
anomalies using the Witten-index of supersymmetric quantum mechanics \ct{alvG},
is to represent $\gam_5$ by $(-1)^F$ where $F$ is a suitably defined fermion
number in the supersymmetric quantum-mechanical model. In the context of the
world-line formulation of spinning particles the latter approach has been
advocated in \ct{mon2}. In this paper I review in some detail the differences
between these approaches and their consequences. I show, that the Witten-index
approach is equivalent to the bosonic version of $\gam_5$ presented in
\ct{jw4},
and corresponds to an irreducible representation of spinors in four
dimensions\footnote{Similar results actually hold also in other dimensions},
whereas the Grassmann-odd representation of $\gam_5$ introduces a doubling of
the number of degrees of freedom, corresponding to a reducible representation
of the Clifford algebra of Dirac matrices used to define spinors.
\vs

\nit
{\bf Dirac equation and supersymmetry}
\vs

The use of supersymmetry in describing spinning particles results from the
formal similarity between the algebra of Dirac operators and the supersymmetry
algebra

\be
Q^2\, =\, H,
\label{1.1}
\ee

\nit
where $Q$ denotes the supercharge and $H$ the hamiltonian operator of a
supersymmetric quantum theory. The Dirac operator $\hat{p} = \gam \cdot p$
is related similarly to the laplacian (the kinetic operator) by

\be
\hat{p}^2\, =\, p_{\mu}^2.
\label{1.2}
\ee

\nit
In the pseudo-classical theory of a massless fermion the classical
quantity corresponding to $\hat{p}$ indeed generates a world-line
supersymmetry, just like the classical hamiltonian generates proper-time
translations.

For massive fermions the problem is a little more complicated: the
Dirac operator $\hat{p} + m$ does not square to the Klein-Gordon operator
$p_{\mu}^2 + m^2$. However, the situation can be saved by introducing
$\gam_5$ and associate

\be
Q\, \rightarrow\, \lh -i \hat{p} + m \rh \gam_5, \hspace{3em}
H\, \rightarrow\, p_{\mu}^2 + m^2.
\label{1.3}
\ee

\nit
This defines a representation of the supersymmetry algebra (\ref{1.1})
owing to the crucial anti-commutation property of the $\gam_5$ operator:

\be
\hat{p} \gam_5 + \gam_5 \hat{p} = 0.
\label{1.4}
\ee

\nit
Clearly the limit $m \rightarrow 0$ is well-defined and defines an alternative
realization of the supersymmetry relation for massless fermions by
$Q = - i \hat{p} \gam_5$.
\vs

\nit
{\bf Grassmann representation of spinors}
\vs

In view of the above, I represent the Dirac algebra in the following by
the pseudo-vector elements $\ps_{\mu} = - i \gam_{\mu} \gam_5$, which obey
the same (Euclidean) anti-commutation relations

\be
\left\{ \ps_{\mu}, \ps_{\nu} \right\}\, =\, 2 \del_{\mu\nu}.
\label{2.1}
\ee

\nit
Notice that one has the relation

\be
\ps_5\, =\, \gam_5\, =\,
  \frac{1}{4!}\, \ve^{\mu\nu\kg\lb} \ps_{\mu} \ps_{\nu} \ps_{\kg} \ps_{\lb},
\label{2.2}
\ee

\nit
with the algebraic properties

\be
\ps_5^2 = 1, \hspace{3em}
\ps_5 \ps_{\mu} + \ps_{\mu} \ps_5 = 0.
\label{1.7}
\ee

\nit
To obtain a pseudo-classical action for fermions one can introduce another
representation of the Dirac algebra in terms of two Grassmann-odd variables
$\xi^{1,2}$ as follows:

\be
\ba{ll}
\dsp{ \ps_1 = \xi^1 + \dd{}{\xi^1}, } &
\dsp{ \ps_2 = -i \lh \xi^1 - \dd{}{\xi^1} \rh, } \\
  & \\
\dsp{ \ps_3 = \xi^2 + \dd{}{\xi^2}, } &
\dsp{ \ps_4 = - i \ps_0 = i \lh \xi^2 - \dd{}{\xi^2} \rh. }
\ea
\label{1.5}
\ee

\nit
It is straightforward to check, that these Grassmann-odd differential operators
satisfy the Clifford algebra (\ref{2.1}) and provide an alternative to the
Dirac matrices in describing spinors.

This representation of the Dirac algebra in terms of anti-commuting variables
can be extended with an element $\ps_5$ with the algebraic properties
(\ref{1.7}) in 2 different ways: first, one can introduce a third Grassmann-odd
variable $\xi^3$ and define

\be
\ps_5 = \xi^3 + \dd{}{\xi^3}, \hspace{3em}
\ps_6 = -i \lh \xi^3 - \dd{}{\xi^3} \rh.
\label{1.8}
\ee

\nit
With these definitions one finds

\be
\left\{ \ps_M, \ps_N \right\}\, =\, 2 \del_{MN}, \hspace{3em}
(M,N) = 1,...,6.
\label{1.9}
\ee

\nit
Therefore both $\ps_5$ and $\ps_6$ satisfy the properties (\ref{1.7}),
and one may arbitrarily chose one of them, say $\ps_5$. The other one then
comes for free, but has no obvious use in the representation of spinors in
4-dimensional space-time at this point (however, see below).

The second way to represent $\ps_5$ is by the non-linear Grassmann-even
expression

\be
\ps_5\, =\,
  \frac{1}{4!}\, \ve^{\mu\nu\kg\lb} \ps_{\mu} \ps_{\nu} \ps_{\kg} \ps_{\lb}\,
  =\, \lh 1 - 2 \xi^1 \dd{}{\xi^1} \rh \lh 1 - 2 \xi^2 \dd{}{\xi^2} \rh.
\label{1.10}
\ee

\nit
This definition corresponds directly to the $4 \times 4$ matrix representation
$\ps_5 = \gam_5$ in (\ref{2.2}) and does have the properties (\ref{1.7}), in
spite of $\ps_5$ being Grassmann-even. However, in this case there is no
analogue of $\ps_6$.

Given these two representations of the Dirac algebra, we consider the
properties of spinors, which are simply defined as functions of the
Grassmann-variables $\xi^k$. First, in the minimal representation with
non-linear Grassmann-even $\ps_5$ a spinor has four components:

\be
\Fg \lh \xi^1, \xi^2 \rh\, =\, \fg_2 + \xi^1 \fg_3 - \xi^2 \fg_4 -
    \xi^1 \xi^2 \fg_1.
\label{1.11}
\ee

\nit
The components have been labeled in a somewhat unconventional way, which has
the advantage that the differential operators (\ref{1.5}, \ref{1.10}) act on
the components in the same way as the Dirac matrices on spinors $\fg_{\ag}$
$(\ag = 1,...,4)$ if they are taken in the chiral basis

\be
\ba{l}
\dsp{ \gam_i\, =\, \lh \ba{cc}
                       0 & -i \sg_i \\
                       i \sg_i & 0 \ea \rh, } \\
 \\
\dsp{ \gam_4\, =\, \lh \ba{cc}
                       0 & 1 \\
                       1 & 0 \ea \rh, \hspace{3em}
      \gam_5\, =\, \lh \ba{cc}
                       1 & 0 \\
                       0 & -1 \ea \rh. }
\ea
\label{1.12}
\ee

\nit
Observe that in this representation $\gam_5$ is diagonal, which is
equivalent to the result

\be
\ps_5 \Fg \lh \xi^1, \xi^2 \rh\, =\, \Fg \lh -\xi^1, -\xi^2 \rh.
\label{1.13}
\ee

\nit
It follows, that in this representation the chirality $\pm$ of the spinor
in the 4-$D$ sense is the same as the Grassmann parity: if $\Fg = \Fg_+ +
\Fg_-$ with

\be
\Fg_+ = \fg_2 - \xi^1 \xi^2 \fg_1, \hspace{3em}
\Fg_- = \xi^1 \fg_3 - \xi^2 \fg_4,
\label{1.14}
\ee

\nit
and the Grassmann parity operator is $(-1)^F$, then

\be
\ps_5 \Fg\, =\, (-1)^F \Fg\, =\, \Fg_+ - \Fg_-.
\label{1.15}
\ee

\nit
As the mass-term in the Dirac operator (\ref{1.3}) contains $\gam_5$,
the Dirac equation becomes

\be
\left[  p^{\mu} \ps_{\mu} + m \ps_5 \right]\, \Fg(\xi^1, \xi^2)\, =\, 0.
\label{1.16}
\ee

\nit
It is therefore clear that the mass-term mixes spinor components of opposite
Grassmann parity. We also observe, that the spinors $\Fg(\xi^1,\xi^2)$ can
be of either Dirac or Majorana type. With the charge conjugation operator
represented by

\be
C\, =\, \lh \xi^1 - \dd{}{\xi^1}\rh \lh \xi^2 - \dd{}{\xi^2} \rh,
\label{1.17}
\ee

\nit
a Majorana spinor has the same expansion (\ref{1.11}) as a Dirac spinor, but
with the additional restriction that $\fg_3 = - \fg_2^*$, $\fg_4 = \fg_1^*$;
in terms of the chiral components:

\be
\Fg_-\, =\, -i \sg_2 \Fg_+.
\label{1.18}
\ee

Next we consider the representation if the Dirac algebra in terms of
three anti-commuting variables $\xi^k$, $k = 1,2,3$. In this case a
general spinor is a function

\be
\Fg \lh \xi^1, \xi^2, \xi^3 \rh\, =\, \Fg_1 \lh \xi^1, \xi^2 \rh\, +\,
    \xi^3 \Fg_2 \lh \xi^1, \xi^2 \rh,
\label{1.19}
\ee

\nit
where each of the functions $\Fg_{1,2}(\xi^1, \xi^2)$ is a 4-component
spinor of the type discussed above. The Dirac equation:

\be
\left[  p^{\mu} \ps_{\mu} + m \ps_5 \right]\, \Fg(\xi^1, \xi^2\, \xi^3)\, =\,
0,
\label{1.20}
\ee

\nit
is now equivalent to an $8 \times 8$ matrix equation

\be
\lh \Gam_{\mu} p^{\mu} + m \Gam_5 \rh \Fg\, =\,
  \lh \ba{cc}
      -i \hat{p} \gam_5 & m \\
       m & i \hat{p} \gam_5 \ea \rh\, \lh \ba{c}
                                          \Fg_1 \\ \Fg_2 \ea \rh\, =\, 0.
\label{1.21}
\ee

\nit
Clearly $(\Gam_{\mu}, \Gam_5)$ define a {\em reducible} 8-dimensional
representation of the 4-$D$ Dirac algebra, as is obtained by dimensional
reduction from six space-time dimensions. This also explains the appearance
of $\ps_6$ in eq.(\ref{1.8}). The four-component spinors $\Fg_{1,2}$ then
describe a degenerate pair of fermions in four space-time dimensions.

The conclusion from this analysis is, that the minimal (Dirac or Majorana)
representation of spinors by Grassmann variables is the four-component
one with a Grassmann-even non-linear representation of $\ps_5$, whilst
the anti-commuting variable representation with Grassmann-odd $\ps_5$
describes a reducible representation of the Dirac algebra with a doubling
of the number of physical degrees of freedom.
\vs

\nit
{\bf Ordered symbols}
\vs

A general linear operator $A$ on a two-component vector $v = (v_1, v_2)$
with matrix

\be
A\, =\, \lh \ba{cc}
            a_0 & a_2 \\
            a_1 & a_3 \ea \rh
\label{3.1}
\ee

\nit
can be represented as an ordered differential operator acting on a function
$v(\xi) = v_1 + \xi v_2$ of an anti-commuting variable $\xi$, with ordered
expansion

\be
A\, =\, a_0 + a_1 \xi + a_2 \dd{}{\xi} + (a_3 - a_0) \xi \dd{}{\xi}.
\label{3.2}
\ee

\nit
The action of such an ordered differential operator is equivalent to an
integral operator

\be
\left[ A v \right] (\xi)\, =\, \int d\xi_1 d\bar{\xi}_1\,
  e^{\bar{\xi}_1 \lh \xi_1 - \xi \rh}\, \bar{A} \lh \xi, \bar{\xi}_1 \rh\,
  v(\xi_1),
\label{3.3}
\ee

\nit
where $\bar{A}(\xi, \bar{\xi})$ is the ordered symbol \ct{ber2,Fad}, defined
as

\be
\bar{A} \lh \xi, \bar{\xi} \rh\, =\, a_0 + a_1 \xi + a_2 \bar{\xi}
        + (a_3 - a_0) \xi \bar{\xi},
\label{3.4}
\ee

\nit
the ordered operator with every derivative $\pl/\pl\xi$ replaced
by the Grassmann variable $\bar{\xi}$. Clearly there is a one-to-one
correspondence between linear operators and ordered symbols. Note that
the unit operator has components $a_0 = a_3 = 1$, $a_1 = a_2 = 0$, and
its symbol is the number 1.

The product of two operators $A$ and $B$ is represented by the symbol

\be
 \bor{\left[ {AB} \right]} \lh \xi, \bar{\xi} \rh\, =\,
  \int d\xi_1\, d\bar{\xi}_1 e^{\lh \bar{\xi}_1 - \bar{\xi} \rh
  \lh \xi_1 - \xi \rh}\, \bar{A} \lh \xi, \bar{\xi}_1 \rh
  \bar{B} \lh \xi_1, \bar{\xi} \rh.
\label{3.5}
\ee

\nit
It is straighforward to write down the ordered symbols and their
multiplication rules for operators on functions of multiple Grassmann
variables. In particular, the symbols for the operators $\ps_{\mu}$
in eq.(\ref{1.5}) take the simple form

\be
\ba{ll}
\dsp{ \bar{\ps}_1 = \xi^1 + \bar{\xi}^1, } &
\dsp{ \bar{\ps}_2 = -i \lh \xi^1 - \bar{\xi}^1 \rh, } \\
  & \\
\dsp{ \bar{\ps}_3 = \xi^2 + \bar{\xi}^2, } &
\dsp{ \bar{\ps}_4 = - i \bar{\ps}_0 = i \lh \xi^2 - \bar{\xi}^2 \rh. }
\ea
\label{3.6}
\ee

\nit
For the symbol of $\ps_5$ we again have two choices: one is to introduce
Grassmann variables $(\xi^3, \bar{\xi}^3)$ and define

\be
\bar{\ps}_5\, =\, \xi^3 + \bar{\xi}^3,  \hspace{3em}
\bar{\ps}_6\, =\, -i \lh \xi^3 - \bar{\xi}^3 \rh,
\label{3.7}
\ee

\nit
for the reducible spinor representation. The other one is to define

\be
\bar{\ps}_5\, =\, \lh 1 - 2 \xi^1 \bar{\xi}^1 \rh
                  \lh 1 - 2 \xi^2 \bar{\xi}^2 \rh\,
              =\, e^{-2 \xi \cdot \bar{\xi}}.
\label{3.8}
\ee

\nit
This is the correct form for the irreducible bosonic representation
of $\ps_5$ corresponding to the Grassmann parity operator $(-1)^F$ \ct{alvG}.
\vs

\nit
{\bf From fields to point particles}
\vs

The quantum theory of (free) fields has an interpretation in terms of
the propagation and exchange of point particles. As emphasized by Schwinger
\ct{Schwing} external sources $J(x)$ for the fields can excite the vacuum,
and the amplitude for exchanging quanta of energy-momentum, spin etc.\ are
proportional to

\be
Z\left[ J \right]\, \sim\, e^{\frac{i}{2}\, \int \bar{J} \cdot G_F \cdot J},
\label{4.1}
\ee

\nit
where $G_F(x-y)$ is the Feynman propagator. This propagator has a direct
interpretation in terms of classical point particles through the
path-integral formalism.\footnote{For a recent discussion see \ct{jw4} and
references therein.} The connection is made as follows: for the particular
case of free fermions, the Feynman propagator in momentum space is

\be
G_F(p)\, =\, \frac{1}{i \hat{p} + m - i \ve}.
\label{4.2}
\ee

\nit
Here the $i\ve$-prescription guarantees the causal behaviour of the theory.
In view of our previous discussion it is actually preferable to work
with the equivalent quantity

\be
G_F(p) \gam_5\, =\, \frac{1}{-i \hat{p} \gam_5 + (m - i\ve) \gam_5}.
\label{4.3}
\ee

\nit
This operator can be represented by its symbol in terms of Grassmann variables
$(\xi^k, \bar{\xi}^k)$ by

\be
\bor{\left[G_F \gam_5\right]}(p;\xi,\bar{\xi})\, =\,
  \frac{p \cdot \bar{\ps}  + m \bar{\ps}_5}{p^2 + m^2}.
\label{4.4}
\ee

\nit
This expression can be rewritten using a generalized Schwinger
representation with a pair of bosonic and fermionic super proper-time
parameters $(T, \sg)$ \ct{fradgit}:

\be
\bor{\left[G_F \gam_5\right]}(p;\xi, \bar{\xi})\, =\, -\frac{i}{2m}\,
  \int_0^{\infty} dT \int d\sg\, e^{-\frac{iT}{2m}(p^2 + m^2) -
  \sg (p \cdot \bar{\ps} + m \bar{\ps}_5)}
\label{4.5}
\ee

\nit
To make the whole exponent proportional to the proper time $T$, so as to
obtain additivity in this parameter, it is customary to define a new
anti-commuting parameter $\chi = m\sg/T$ and write

\be
\bor{\left[G_F \gam_5\right]}(p;\xi, \bar{\xi})\, =\, -\frac{i}{2}\,
 \int_0^{\infty} \frac{dT}{T}\, \int d\chi\, \tK(p; \xi, \bar{\xi}; T, \chi),
\label{4.6}
\ee

\nit
where the integrand is

\be
\tK(p; \xi, \bar{\xi}; T, \chi)\, =\, e^{-\frac{iT}{2m}\, \left[ p^2 +
  m^2 - i\ve - 2 i \chi (p \cdot \bar{\ps} + m \bar{\ps}_5) \right]}
\label{4.7}
\ee

\nit
The integral kernel may be interpreted as the Fourier transform of the
real-space expression

\be
\ba{lll}
K(x-y; \xi, \bar{\xi}; T, \chi) & = & \dsp{ \int \frac{d^4p}{(2\pi)^4}\,
       e^{ip \cdot (x-y)}\, \tK(p;\xi, \bar{\xi}; T, \chi) } \\
  & & \\
  & = & \dsp{ -i \lh \frac{m}{2\pi T} \rh^2\, e^{\frac{im}{2T}\, (x-y)^2 -
        \frac{iT}{2}\, (m-i\ve) - \chi \bar{\ps} \cdot (x-y) - T \chi
        \bar{\ps}_5}. }
\ea
\label{4.8}
\ee

\nit
The integral kernel (\ref{4.8}) has two useful properties:\nl
A.\ in the limit $T \rightarrow 0$ it reduces to the unit distribution

\be
\lim_{T \rightarrow 0} K(x-y; \xi, \bar{\xi}; T, \chi)\, =\,
\del^4 \lh x-y \rh;
\label{4.9}
\ee

\nit
B.\ it satisfies the composition rule

\be
\ba{l} \dsp{
\int d^4z \int \prod_k \left[ d\xi^{\prime\, k} d\bar{\xi}^{\prime k} \right]\,
     e^{\lh \pr{\bar{\xi}} - \bar{\xi} \rh \cdot \lh \pr{\xi} - \xi \rh}\,
     K(x-z; \xi, \bar{\xi}^{\prime}; \pr{T}, \chi)
     K(z-y; \xi^{\prime}, \bar{\xi}; T^{\prime\prime}, \chi)\, =\, } \\
   \\
     \hspace{4em}
     =\, K(x-y; \xi, \bar{\xi}; T^{\prime} + T^{\prime\prime}, \chi).
\ea
\label{4.10}
\ee

\nit
Note that all results so far are true for both representations of $\ps_5$.
The properties A and B permit a path-integral representation of the
propagator, by iterating the composition rule $N$ times for periods
$\Del T = T/(N+1)$ and taking the limit $N \rightarrow \infty$:

\be
\ba{lll}
K(x-y; \xi, \bar{\xi}; T\, \chi) & = & \dsp{ \int \prod_{k=1}^N \left[
 d^4x_k d^n\xi_k d^n\bar{\xi}_k \right]\, e^{\frac{1}{2}\, \sum_{j=1}^N \left[
 \lh \bar{\xi}_j - \bar{\xi}_{j-1}\rh \cdot \xi_j - \bar{\xi}_j \cdot
 \lh \xi_{j+1} - \xi_j \rh \right]} } \\
  & & \\
  & \times & \dsp{ e^{\frac{1}{2}\, \lh \bar{\xi}_0 - \bar{\xi}_N \rh \cdot
    \xi_{N+1} - \frac{1}{2}\, \bar{\xi}_0 \cdot \lh \xi_1 - \xi_{N+1} \rh}\,
    \prod_{s=0}^N K(x_{s+1} - x_s; \xi_{s+1}, \bar{\xi}_s; \Del T,\chi).
  }\\
  & & \\
  & \rightarrow & \dsp{ \int_y^x \cD^4 x(\tau)
    \int^{\xi}\int_{\bar{\xi}} \cD^n \xi(\tau) \cD^n \bar{\xi}(\tau)\,
    e^{i S_T\left[ x(\tau), \xi(\tau), \bar{\xi}\,(\tau) \right]}. }
\ea
\label{4.11}
\ee

\nit
Here I have labeled $x_0 = y$, $x_{N+1} = x$, $\bar{\xi}_0 = \bar{\xi}$ and
$\xi_{N+1} = \xi$, and $n = 2$ or 3 depending on the representation of
$\ps_5$. In the continuum limit the classical action in the exponent
takes the form

\be
S_T\left[ x(\tau), \xi(\tau), \bar{\xi}(\tau) \right]\, =\,
 \int_0^T d\tau\, \left[ \frac{m}{2}\, \lh \dot{x}_{\mu}^2 - 1\rh -
 \frac{i}{2}\,  \lh \dot{\bar{\xi}} \cdot \xi - \bar{\xi} \cdot \dot{\xi} \rh
 + i \chi\, \bar{\ps} \cdot \dot{x} + i \chi\, \bar{\ps}_5 \right],
\label{4.12}
\ee

\nit
modulo boundary terms. This effective action defines the pseudo-classical
theory in correspondence with the quantum field theory of fermions described
by the propagator $G_F$. It is well-known \ct{alvG} that for closed
propagators (loops) with anti-periodic fermionic boundary conditions:
$\xi_k = - \xi_{N+k}$, $\bar{\xi}_k = - \bar{\xi}_{N+k}$, the boundary terms
disappear. The resulting path-integral expression then corresponds to the
trace of $\gam_5$ times the Dirac propagator of the field-theory:

\be
\Tr \left[\gam_5 \lh \hat{\pl} + m \rh^{-1} \right]\, =\,
    -\frac{i}{2}\, \int_0^{\infty} \frac{dT}{T}\, \int d\chi\,
    \oint_{PBC} \cD^4 x(\tau) \oint_{ABC} \cD^n \xi(\tau) \cD^n
\bar{\xi}(\tau)\,
    e^{i S_T\left[ x(\tau), \xi(\tau), \bar{\xi}(\tau) \right]},
\label{4.13}
\ee

\nit
where $PBC$ and $ABC$ denote periodic and anti-periodic boundary conditions
respectively.

For the case of the bosonic realization $\ps_5 = (-1)^F$ (indicated by +)
the trace of $(\hat{\pl} + m)^{-1}$ itself is then obtained from the same
path-integral with {\em periodic} boundary conditions:

\be
\ba{lll}
\Tr\, G_F^{(+)} & = & \dsp{ \dd{}{m} \log \det \lh \hat{\pl} + m \rh } \\
  & &  \\
  & = & \dsp{ -\frac{i}{2}\, \int_0^{\infty} \frac{dT}{T}\, \int d\chi\,
  \oint_{PBC} \cD^4 x(\tau) \oint_{PBC} \cD^2 \xi(\tau) \cD^2 \bar{\xi}(\tau)\,
  e^{i S_T^{(+)}\left[ x(\tau), \xi(\tau), \bar{\xi}(\tau) \right]}. }
\ea
\label{4.13.1}
\ee

\nit
Note that this is in contrast to the formulation of \ct{alvG}, where the trace
of the Dirac operator is obtained with anti-periodic boundary conditions.
However, integration over $\chi$ brings down another $\bar{\ps}_5$ which
changes the boundary conditions back to anti-periodic:

\be
\Tr\, G_F^{(+)}\, =\,
  \frac{i}{2}\, \int_0^{\infty} dT \oint_{PBC} \cD^4 x(\tau)
  \oint_{ABC} \cD^2 \xi(\tau) \cD^2 \bar{\xi}(\tau)\, \ld
  e^{i S_T^{(+)}\left[ x(\tau), \xi(\tau), \bar{\xi}(\tau) \right]}
  \right|_{\chi = 0}.
\label{4.13.2}
\ee

\nit
For the case of the linear fermionic realization of $\ps_5$ (indicated by $-$)
a direct calculation shows that after integration over $(\xi^3,
\bar{\xi}^3)$ and $\chi$ the trace of the propagator can be written as

\be
\Tr\, G_F^{(-)}\, =\, i \int_0^{\infty} dT \oint_{PBC} \cD^4 x(\tau)
    \oint_{ABC} \cD^2 \xi(\tau) \cD^2 \bar{\xi}(\tau)\,
    \ld e^{i S_T^{(-)}\left[ x(\tau), \xi(\tau), \bar{\xi}(\tau) \right]}
    \right|_{\chi = \xi^3 = \bar{\xi}^3 = 0},
\label{4.14}
\ee

\nit
with only an integration over the paths for the remaining two components of
$(\xi(\tau), \bar{\xi}(\tau))$, satisfying anti-periodic boundary conditions,
left. The results (\ref{4.13.2}) and (\ref{4.14}) are the same for both
prescriptions for dealing with $\ps_5$, up to a factor of 2. This is precisely
the factor one expects from doubling the dimension of spinor space, as I have
argued to be associated with the Grassmann-odd representation of $\ps_5$.
\vs

\nit
{\bf Supersymmetry}
\vs

In terms of the covariant fermionic co-ordinates $\bar{\ps}_{\mu}$,
eq.(\ref{3.6}), the classical action (\ref{4.12}) for $\chi = 0$ (and
in the appropriate case $\xi^3 = \bar{\xi}^3 = 0$) becomes

\be
S_T\left[ x(\tau), \bar{\ps}(\tau) \right]_{\chi = 0}\, =\,
 \int_0^T d\tau\, \left[ \frac{m}{2}\, \lh \dot{x}_{\mu}^2 - 1\rh
 + \frac{i}{4}\, \bar{\ps} \cdot \dot{\bar{\ps}} \right].
\label{5.1}
\ee

\nit
Both path-integrals (\ref{4.13.2}) and (\ref{4.14}) therefore admit a
rigid supersymmetry transforming $x^{\mu}$ into $\bar{\ps}^{\mu}$

\be
\del x^{\mu} = \bar{\ps}^{\mu} \ve, \hspace{3em}
\del \bar{\ps}^{\mu} = 2m \dot{x}^{\mu} \ve.
\label{5.2}
\ee

\nit
This is of course to be expected from the algebraic starting point
(\ref{1.3}). However, in the full action (\ref{4.12}) with $\chi$ reinstated
only the action with fermionic $\bar{\ps}_5$ can be interpreted as the
gauge-fixed version of a model with local world-line supersymmetry, where
the einbein field is fixed to the constant value $T$ and the gravitino
fixed to the Grassmann constant $\chi$ \ct{jw4}. In this formulation there
is apparently no matching between the four bosonic and the six fermionic
degrees of freedom, but this is because $\bar{\ps}_{5,6}$ are members
of a fermionic supermultiplet of which the bosonic component is an
auxiliary field (a recent review of $D=1$ supergravity can be found in
\ct{jw3}). The additional fermion variable $\bar{\ps}_6$ enters the
action in the path integral through the kinetic terms; however, in the
free-particle theory it is completely decoupled from the rest of the physical
degrees of freedom and actually describes a simple topological quantum theory
\ct{jw4}.

In contrast, in the case of a Grassmann-even $\bar{\ps}_5$ local
supersymmetry of the action $S_T$ seems violated by the Grassmann-odd term
$i \chi \bar{\ps}_5$, which is rather unusual in a classical Lagrangian
formalism. However, the presence of such a term is well-understood as
the fermion mass-term flips chirality, which in the present formalism is
equivalent to a change in Grassmann parity of the state. \nl
\vs

\nit
{\bf Acknowledgement}
\vs

\nit
I wish to thank Michael Schmidt for an invitation to visit the Institute for
Theoretical Physics in Heidelberg, and both him and Martin Reuter for useful
discussions about the topic of this paper.

\end{document}